\documentclass[prd,twocolumn,amsmath,amssymb,nofootinbib,floatfix,superscriptaddress]{revtex4}

\usepackage{graphicx,bm,float,physics}
\usepackage{xcolor}
\usepackage[normalem]{ulem}

\makeatletter
\def\graphicscale{\twocolumn@sw{0.3}{0.4}}
\def\graphicthreescale{\twocolumn@sw{0.3}{0.4}}

\begin{document}

\title{Charged critical behavior and nonperturbative continuum
  limit\\ of three-dimensional lattice SU($N_c$) gauge Higgs models}

\author{Claudio Bonati} 
\affiliation{Dipartimento di Fisica dell'Universit\`a di Pisa and 
INFN Sezione di Pisa, Largo Pontecorvo 3, I-56127 Pisa, Italy}

\author{Andrea Pelissetto}
\affiliation{Dipartimento di Fisica dell'Universit\`a di Roma Sapienza
and INFN, Sezione di Roma I, I-00185 Roma, Italy}

\author{Ivan Soler Calero} 
\affiliation{Dipartimento di Fisica dell'Universit\`a di Pisa and 
INFN Sezione di Pisa, Largo Pontecorvo 3, I-56127 Pisa, Italy}

\author{Ettore Vicari} 
\affiliation{Dipartimento di Fisica dell'Universit\`a di Pisa,
        Largo Pontecorvo 3, I-56127 Pisa, Italy}

\date{\today}

\begin{abstract}
  We consider three-dimensional (3D) lattice SU($N_c$) gauge Higgs
  theories with multicomponent ($N_f>1$) degenerate scalar fields and
  U($N_f$) global symmetry, focusing on systems with $N_c=2$, to
  identify critical behaviors that can be effectively described by the
  corresponding 3D SU($N_c$) gauge Higgs field theory.  The
  field-theoretical analysis of the RG flow allows one to identify a
  stable charged fixed point for large values of $N_f$, that would
  control transitions characterized by the global symmetry-breaking
  pattern ${\rm U}(N_f)\rightarrow \mathrm{SU}(2)\otimes
  \mathrm{U}(N_f-2)$.  Continuous transitions with the same
  symmetry-breaking pattern are observed in the SU(2) lattice gauge
  model for $N_f \ge 30$.  Here we present a detailed finite-size
  scaling analysis of the Monte Carlo data for several large values of
  $N_f$.  The results are in substantial agreement with the
  field-theoretical predictions obtained in the large-$N_f$
  limit. This provides evidence that the SU($N_c$) gauge Higgs field
  theories provide the correct effective description of the 3D
  large-$N_f$ continuous transitions between the disordered and the
  Higgs phase, where the flavor symmetry breaks to
  $\mathrm{SU}(2)\otimes \mathrm{U}(N_f-2)$.  Therefore, at least for
  large enough $N_f$, the 3D SU($N_c$) gauge Higgs field theories with
  multicomponent scalar fields can be nonperturbatively defined by the
  continuum limit of lattice discretizatized models with the same
  local and global symmetries.

\end{abstract}

\maketitle

% ========================= BODY =========================

\section{Introduction}
\label{intro}

Local gauge symmetries play a fundamental role in the construction of
quantum and statistical field theories that describe phenomena in
various physical contexts: In high-energy physics they are used to
formulate the theories of fundamental
interactions~\cite{Weinberg-book1, Weinberg-book2, ZJ-book,
  Georgi-book}, in condensed-matter physics their application spans
from superconductors to systems with topologically ordered
phases~\cite{Anderson-book,Wen-book}, in statistical mechanics they
are needed to describe classical and quantum critical phenomena with
(also emergent) gauge fields~\cite{Sachdev-19}.

The physical properties of lattice gauge models with scalar fields
crucially depend on the behavior of gauge and scalar
modes~\cite{GG-72, HLM-74, OS-78, FS-79, DRS-80, Hikami-80, BN-87,
  SSSNH-02, MZ-03, NRR-03, SBSVF-04, DP-14, PV-19-AH3d, SSST-19,
  BPV-19, BPV-20, SPSS-20, BPV-20-on, BPV-21, BFPV-21-su-ad,
  BFPV-21-su, BPV-22, BPV-23, BPV-24}.  Their interplay can give rise
to continuous phase transitions, which are associated with nontrivial
continuum limits of the corresponding gauge theories. The
corresponding critical behavior depends both on the breaking pattern
of the global symmetry and on the local gauge symmetry, which
determines which scalar degrees of freedom can become
critical. Moreover, in the presence of gauge symmetries, scalar
systems show Higgs phases~\cite{Anderson-63,SSBgauge}, a fundamental
feature of many modern-physics systems.

In this paper we focus on a class of three-dimensional (3D)
non-Abelian Higgs (NAH) field theories, which are characterized by
SU($N_c$) gauge invariance and by the presence of $N_f$ degenerate
scalar fields transforming in the fundamental representation of the
gauge group. The fundamental fields a complex scalar field
$\Phi^{af}(\bm{x})$, where $a=1,...,N_c$ and $f=1,\ldots,N_f$, and a
gauge field $A_{\mu}^c(\bm{x})$, where $c=1,\ldots,N_c^2-1$.  The most
general renormalizable Lagrangian consistent with the local SU($N_c$)
{\em color} symmetry and the global U($N_f$) {\em flavor} symmetry of
the scalar sector is
\begin{equation} \label{cogau} 
\begin{aligned}
{\cal L}=& \frac{1}{g^2} {\rm Tr}\,F_{\mu\nu}^2 + {\rm Tr} \, [(D_\mu
  \Phi)^\dagger (D_\mu \Phi)]\\
  +& r\,{\rm Tr}\,\Phi^\dagger\Phi 
  + \,{u\over 4} \,({\rm Tr}\,\Phi^\dagger\Phi)^2 \,+ \,{v\over 4}\,
  {\rm Tr}\,(\Phi^\dagger\Phi)^2 \,,
\end{aligned}
\end{equation}
where $F_{\mu\nu} = \partial_\mu A_\nu -\partial_\nu A_\mu -i[A_\mu,
  A_\nu]$ (with $A_{\mu, ab}=A_{\mu}^c t_{ab}^c$), and $D_{\mu, ab} =
\partial_\mu\delta_{ab} -i t_{ab}^c A_\mu^c$, where $t^c_{ab}$ are the
SU($N_c$) Hermitian generators in the fundamental representation.

The Lagrangian \eqref{cogau} has been written in the standard
continuum form, in which perturbative computations are usually carried
out (after gauge fixing). An important issue is whether it is possible
to give a definition of the model that goes beyond perturbation
theory.  To investigate this issue, one may proceed as it is usually
done in quantum chromodynamics (QCD), where the question is studied by
considering the lattice QCD formulation~\cite{Wilson-74,MM-book}.  In
this setting a nonperturbative continuum limit exists if the lattice
regularized model undergoes a continuous transition with a divergent
length scale, in which all fields become critical.

Thus, the crucial point is the identification of critical transitions
in 3D lattice NAH models.  In the field-theoretical setting this is
equivalent to the existence of a stable fixed point (FP) of the
renormalization-group (RG) flow of the 3D NAH field theory
(\ref{cogau}).  Its existence allows us to define a continuum limit
and therefore it would provide a nonperturbative definition of the
model, as it occurs in the case of QCD~\cite{Wilson-74,MM-book}.

This program has been carried out in 3D Abelian Higgs (AH) theories
(scalar electrodynamics).  Noncompact lattice formulations of the U(1)
gauge fields~\cite{BPV-21}, and compact formulations with
higher-charge scalar fields~\cite{BPV-20} undergo continuous
transitions, where scalar and gauge modes become critical, allowing us
to define a corresponding scalar-gauge statistical field theory.  Note
that the identification of the correct nonperturbative continuum limit
is not trivial, since 3D lattice AH models also undergo continuous
transitions that are not related with the gauge field theory. Indeed,
there are transitions where gauge modes play no role and that have an
effective Landau-Ginzburg-Wilson (LGW) description with no local gauge
symmetry~\cite{BPV-21,BPV-22}, and topological transitions only driven
by the gauge fields, where scalar fields play no role~\cite{BPV-24}.
None of these transitions, even if continuous, allows one to define
the continuum limit of the gauge Higgs field theory, which requires
both gauge and scalar modes to be critical.

For this reason, in order to correctly identify the continuous
transitions that provide the continuum limit for the corresponding
field theory, it is crucial to compare the lattice results with an
independent calculation. In the case of the lattice AH models, the
identification was supported by the comparison of the numerical
lattice results with nonperturbative field-theoretical computations in
the limit of a large number of components of the scalar
field~\cite{BPV-20,BPV-21,BPV-22,BPV-24}.

In this paper, we wish to pursue the same program for the NAH field
theory (\ref{cogau}). The RG flow in the space of the Lagrangian
couplings has been analyzed to one-loop order~\cite{Hikami-80}, close
to four dimensions, in the $\varepsilon\equiv 4-d$
expansion~\cite{WK-74}. It has a stable infrared FP, with positive
quartic coupling $v$, for any $N_c$ and sufficiently large $N_f$
~\cite{BFPV-21-su}.  We qualify this FP as {\em charged}, because the
gauge coupling assumes a nonzero positive value, thus implying
nontrivial critical correlations of the gauge field. These one-loop
$\varepsilon$-expansion results only indicate that a continuum limit
can be defined for large $N_f$, but do not provide a quantitative
characterization of the behavior in three dimensions. Thus, they do
not provide quantitative results that can be compared with numerical
estimates obtained in the corresponding three-dimensional lattice
model. For this purpose the nonperturbatice large-$N_f$ expansion at
fixed $N_c$ is more useful: $O(1/N_f)$ estimates of critical exponents
~\cite{Hikami-80} can be used to verify the correspondence of lattice
results and field-theory estimates.

In this work we mostly focus on lattice NAH models with SU(2) gauge
symmetry. Their phase diagram was investigated in
Ref.~\cite{BFPV-21-su}, identifying different transition lines.  In
this paper we report a numerical study of some of these transitions,
with the purpose of verifying if the observed critical behavior is
consistent with the predictions of the NAH field theory.  We perform
Monte Carlo (MC) simulations for sufficiently large $N_f$, and
finite-size scaling (FSS) analyses of the MC results to estimate the
universal features of the transitions.  The numerical estimates of the
$N_f$-dependent critical exponents are then compared with the results
obtained by using the $1/N_f$ field-theoretical
expansion~\cite{Hikami-80}.  Our numerical results for the
length-scale exponent $\nu$ agree with the $1/N_f$ prediction,
providing a robust evidence of the fact that the lattice NAH models
develop critical behaviors that can be associated with the stable
charged FP of the RG flow of the NAH field theory.

It is worth emphasizing that the existence of these new universality
classes -- characterized by the presence of a non-Abelian gauge
symmetry -- not only establish the nonperturbative existence of a new
class of 3D quantum field theories, but also allow us to extend the
phenomenology of continuous transitions of 3+1 dimensional lattice
gauge theories at finite temperature, see, e.g., Refs.~\cite{PW-84,
  Nadkarni:1989na, Kajantie:1993ag, Buchmuller:1994qy, AY-94,
  Laine-95, Kajantie:1996mn, Meyer-Ortmanns:1996ioo, Hart:1996ac,
  BPV-03, BVS-06, PV-13}.

The paper is organized as follows. In Sec.~\ref{sft} we collect the
known results on the RG flow of the NAH field theory (\ref{cogau}),
based on $\varepsilon$ expansion, and the large-$N_f$ nonperturbative
predictions.  In Sec.~\ref{model} we define the lattice NAH models,
essentially obtained by discretizing the NAH field theory, and discuss
some general features of their phase diagram.  In Sec.~\ref{numres} we
present the FSS analyses of the numerical MC data obtained for $N_c=2$
and $N_f =30,40,60$.  Finally, we draw our conclusions in
Sec.~\ref{conclu}.

\section{NAH field theory}
\label{sft}

\subsection{RG flow and large-$N_f$ predictions}
\label{sft-rg}

The RG flow of the field theory (\ref{cogau}) was determined close to
four dimensions in the framework of the $\varepsilon\equiv 4-d$
expansion~\cite{WK-74}. The RG functions were computed by using
dimensional regularization and the minimal-subtraction (MS)
renormalization scheme, see, e.g., Ref.~\cite{ZJ-book,PV-02}.  The RG
flow is determined by the $\beta$ functions associated with the
Lagrangian couplings $u$, $v$, and $\alpha = g^2$.  At one-loop order
they are given by ~\cite{Hikami-80,BFPV-21-su}
\begin{eqnarray}
\beta_\alpha &=& -
  \varepsilon \alpha + (N_f-22N_c)\,\alpha^2,
\label{betaalpha}\\
\beta_u &=& -\varepsilon u + (N_f N_c + 4) u^2   
+ 2 (N_f+N_c) u v + 3 v^2  \nonumber \\
&& - {18 \,(N_c^2 -1)\over N_c} \, u \alpha 
+ {27 (N_c^2 + 2)\over N_c^2}\,\alpha^2,
\label{betau}\\
\beta_v &=& - \varepsilon v + (N_f+N_c)v^2 + 6uv
- {18 \,(N_c^2 -1)\over N_c} \, v \alpha
\nonumber\\
&&+ {27 (N_c^2 - 4)\over N_c}\,\alpha^2.
\label{betav}
\end{eqnarray}
Some numerical factors, which can be easily inferred from the above
expressions, have been reabsorbed in the normalizations of the
renormalized couplings to simplify the expressions.

The analysis of the common zeroes of the $\beta$
functions~\cite{BFPV-21-su} shows that the RG flow close to four
dimensions has a stable charged FP with a nonvanishing $\alpha$ if
$N_f > N_f^*$, where $N_f^*$ depends on $N_c$ and on the space
dimension.  Close to four dimensions, we have $N_f^*=
375.4+O(\varepsilon)$ for $N_c=2$, and $N_f^* = 638.9+O(\varepsilon)$
for $N_c=3$. The stable charged FP lies in the region $v>0$ for any
$N_c$.  The number of components $N_f^*$ necessary to have a stable
charged FP is quite large in four dimensions. However, we expect
$N_f^*$ to significantly decrease in three dimensions, as it happens
in the AH theories~\cite{IZMHS-19,BPV-21,BPV-22,BPV-23}, where it
varies from $N_f^*\approx 183$ in four dimensions~\cite{HLM-74} to a
number in the range $4<N_f^*<10$ in three dimensions~\cite{BPV-21}
(see also Refs.~\cite{IZMHS-19,SZJSM-23}).

As we already mentioned in the introduction, the one-loop
$\varepsilon$ expansion provides only qualitative informations for
three dimensional systems.  A more quantitative approch is the 1/$N_f$
expansion at fixed $N_c$ ~\cite{Hikami-80}. Assuming the existence of
a charged critical behavior for finite $N_f$, this approach provides
3D predictions of the critical quantities for large values of
$N_f$. The length-scale critical exponent $\nu$ for was computed to
$O(N_f^{-1})$~\cite{Hikami-80}, obtaining
\begin{eqnarray}
  \nu =  1 - {48 N_c\over \pi^2 N_f} + O(N_f^{-2}),
  \label{nulargen}
\end{eqnarray}
for tree-dimensional systems.  In particular, $\nu \approx 1 - 9.727/N_f$ for
$N_c=2$.

\subsection{Relevance of the field-theoretical results}
\label{sft-relevance}

The studies of the continuous transitions and critical behaviors of
lattice Abelian and non-Abelian gauge theories with scalar matter,
see, e.g.,
Refs.~\cite{BPV-19,PV-19-AH3d,SSST-19,BPV-20,SPSS-20,BPV-20-on,
  BPV-21,BFPV-21-su-ad,BFPV-21-su,BPV-22,BPV-23,BPV-24}, have shown
the emergence of several qualitatively different types of transitions.

In some cases only gauge-invariant scalar-matter correlations become
critical at the transition, while the gauge variables do not display
long-range correlations.  At these transitions, gauge fields prevent
non-gauge invariant scalar correlators from acquiring nonvanishing
vacuum expectation values and from developing long-range order.  In
other words, the gauge symmetry hinders some scalar degrees of
freedom---those that are not gauge invariant---from becoming critical.
In this case the critical behavior or continuum limit is driven by the
condensation of a gauge-invariant scalar order parameter, usually
provided by a gauge-invariant composite operator of the scalar fields.
This composite operator plays the role of fundamental field in the LGW
theory which provides an effective description of the critical
behavior.  Therefore, the effective LGW description depends only on the
scalar order-parameter field, and is only characterized by the global
symmetry of the model.  Gauge invariance is only relevant in
determining the gauge-invariant scalar order parameter.  Examples of
such continuous transitions are found in lattice AH
models~\cite{PV-19-AH3d,BPV-21,BPV-22}, and lattice NAH
models~\cite{BPV-19,BPV-20-on,BFPV-21-su}. A more complex example is
the finite-temperature chiral transitions in QCD.  Ref.~\cite{PW-84}
(see also Refs.~\cite{BPV-03,PV-13}) assumed this transition to be
only driven by the fermionic related modes, proposing an effective LGW
theory in terms of a scalar gauge-invariant composite operator
bilinear in the fermionic fields, without gauge fields.

There are also examples of phase transitions in lattice gauge models
where scalar-matter and gauge-field correlations are both critical. In
this case the critical behavior is expected to be controlled by a
charged FP in the RG flow of the corresponding continuum gauge field
theory. This occurs, for instance, in the 3D lattice AH model with
noncompact gauge fields~\cite{BPV-21,BPV-23}, and in the compact model
with scalar fields with higher charge $Q\ge 2$~\cite{BPV-20}, for a
sufficiently large number of scalar components. Indeed, the critical
behavior along one of their transition lines is associated with the
stable FP of the AH field
theory~\cite{HLM-74,DHMNP-81,FH-96,YKK-96,MZ-03,KS-08,IZMHS-19},
characterized by a nonvanishing gauge coupling.

As already mentioned in the introduction, at present, there is no
conclusive evidence that 3D NAH lattice models undergo continuous
transitions with both scalar and gauge critical correlations, which
can be associated with the stable charged FP of their RG flow
discussed in Sec.~\ref{sft-rg}.  A preliminary study was reported in
Ref.~\cite{BFPV-21-su}.  In this paper we return to this issue,
comparing more accurate numerical analyses with the results obtained
in the field-theoretical 1/$N_f$ expansion. In particular, we
investigate whether, along some specific transition lines, the
critical behavior is characterized by a critical exponent $\nu$ that
is consistent, for large values of $N_f$, with the nonperturbative
$1/N_f$ result~(\ref{nulargen}).

\section{Lattice SU($N_c$) gauge models with multiflavor scalar fields}
\label{model}

\subsection{The lattice model}
\label{lattmod}

As in lattice QCD~\cite{Wilson-74}, we consider lattice SU($N_c$)
gauge models which are lattice discretizations of the NAH field theory
(\ref{cogau}). They are defined on a cubic lattice of linear size $L$
with periodic boundary conditions.  The scalar fields are complex
matrices $\Phi^{af}_{\bm x}$ (with $a=1,...,N_c$ and $f=1,...,N_f$),
satisfying the unit-length constraint ${\rm Tr}\, \Phi_{\bm x}^\dagger
\Phi_{\bm x}^{\phantom\dagger} = 1$, defined on the lattice sites,
while the gauge variables are ${\rm SU}(N_c)$ matrices $U_{{\bm
    x},\mu}$~\cite{Wilson-74} defined on the lattice links.  The
lattice Hamiltonian reads~\cite{BFPV-21-su}
\begin{align}
H =&- J\,N_f\sum_{{\bm x},\mu} {\rm Re}\,{\rm Tr} \,\Phi_{\bm
    x}^\dagger \, U_{{\bm x},\mu} \, \Phi_{{\bm
      x}+\hat{\mu}}^{\phantom\dagger} + {v\over 4} \sum_{\bm x} {\rm
    Tr}\,(\Phi_{\bm x}^\dagger\Phi_{\bm x})^2 \nonumber \\ 
  &-
  {\gamma\over N_c} \sum_{{\bm x},\mu>\nu} {\rm Re} \, {\rm Tr}\,
  [U_{{\bm x},\mu} \,U_{{\bm x}+\hat{\mu},\nu} \,U_{{\bm
        x}+\hat{\nu},\mu}^\dagger \,U_{{\bm x},\nu}^\dagger].
\label{lattNAH}
\end{align}
In the following we set $J=1$, so that energies are measured in units
of $J$, and write the partition function as $Z = \sum_{\{\Phi,U\}}
\exp(-\beta H)$ where $\beta=1/T$.

The Hamiltonian $H$ is invariant under local SU($N_c$) and global
U($N_f$) transformations. Note that U($N_f$) is not a simple group and
thus we may separately consider SU($N_f)$ and U(1) transformations,
that correspond to $\Phi^{af} \to \sum_g V^{fg} \Phi^{ag}$, $V\in$
SU($N_f$), and $\Phi^{af} \to e^{i\alpha} \Phi^{ag}$, $\alpha \in
[0,2\pi)$, respectively.  Since the diagonal matrix with entries
  $e^{2\pi i/N_c}$ is an SU($N_c$) matrix, $\alpha$ can be restricted
  to $[0,2\pi/N_c)$ and the global symmetry group is more precisely
    U$(N_f)/\mathbb{Z}_{N_c}$ when $N_f\ge N_c$ (if $N_f<N_c$ a global
    U(1)/$\mathbb{Z}_{N_c}$ transformation can be reabsorbed by a
    SU($N_c$) gauge transformation, see Ref.~\cite{BPV-19}).

Note that the parameter $v$ in the lattice Hamiltonian corresponds to
the Lagrangian parameter $v$ in Eq.~(\ref{cogau}).  Therefore, if the
lattice model (\ref{lattNAH}) develops a critical behavior described
by the charged FP of the NAH field theory, then this is expected to
occur for positive values of $v$.

\subsection{The phase diagrams for $N_f>N_c$}
\label{phadia}

A thorough discussion of the phase diagram of the lattice NAH models
(\ref{lattNAH}) was reported in Ref.~\cite{BFPV-21-su}. In this
section we recall the main features that are relevant for the present
study. For $N_f=1$ the phase diagram is trivial, as only one phase is
present ~\cite{OS-78,FS-79,DRS-80}.  For $N_f>1$, the lattice model
has different low-temperature Higgs phases, which are essentially
determined by the minima of the scalar potential $v\,{\rm
  Tr}(\Phi^\dagger \Phi)^2$ with the unit-length constraint ${\rm
  Tr}\,\Phi^\dagger \Phi=1$.  Their properties crucially depend on the
sign of the parameter $v$, the number $N_c$ of colors, and the number
$N_f$ of flavors. Substantially different behaviors are found for
$N_f>N_c$, $N_f=N_c$, and $N_f< N_c$. Also $N_c$ is relevant and one
should distinguish systems with $N_c=2$ from those with $N_c\ge 3$.
Since we are interested in phase transitions that can be described by
the stable charged FP of the NAH field theory, and we want to compare
their features with the large-$N_f$ predictions at fixed $N_c$, we
focus on the case $N_f>N_c$.

\begin{figure}[tbp]
\includegraphics[width=0.9\columnwidth, clip]{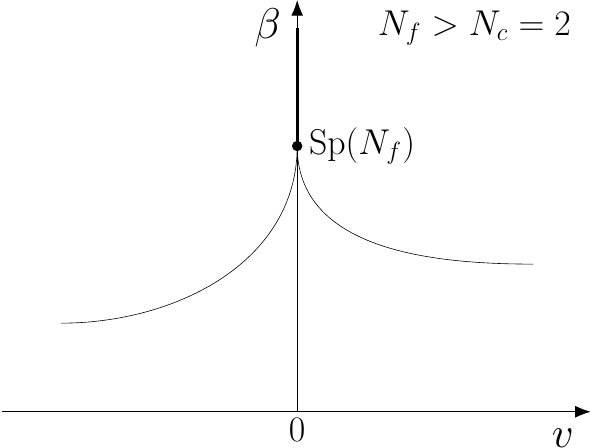}
\caption{A sketch of the phase diagram expected for $N_f>N_c=2$ for
  fixed values of $\gamma\ge 0$. For $v < 0$, $\gamma$ should not play
  any role, and the transition line at fixed $\gamma>0$ is generally
  expected to be of first order. For $v> 0$, the nature of the
  transition might depend on $\gamma$ for sufficiently large values of
  $N_f$. For $v=0$ we have a first-order transition line for large
  $\beta$, ending at a finite value of $\beta$.  See
  Ref.~\cite{BFPV-21-su} for more details.  }
  \label{phdianfgtnc2}
\end{figure}

\begin{figure}[tbp]
\includegraphics[width=0.9\columnwidth, clip]{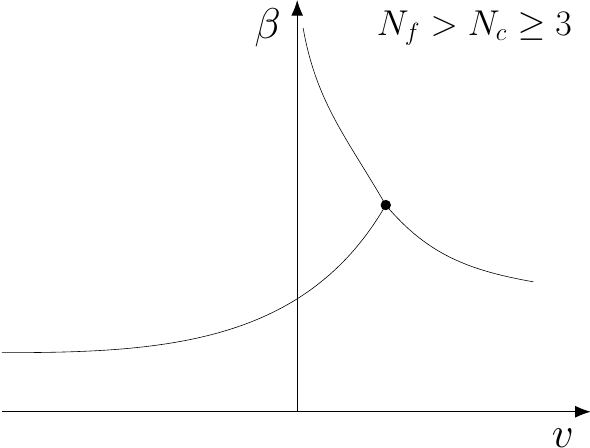}
\caption{A sketch of the phase diagram expected for $N_f>N_c\ge 3$ for
  fixed values of $\gamma\ge 0$. For $v < 0$, $\gamma$ should not play
  any role and the transition line should be generally of first order.
  For $v > 0$, the nature of the transition might depend on $\gamma$
  for sufficiently large values of $N_f$. See Ref.~\cite{BFPV-21-su}
  for more details.  }
  \label{phdianfgtncge3}
\end{figure}

Sketches of the phase diagrams for $N_c=2$ and $N_c\ge 3$ when
$N_f>N_c$ are shown in Figs.~\ref{phdianfgtnc2} and
\ref{phdianfgtncge3}, respectively. They are qualitatively similar,
with two different Higgs phases and a single high-temperature
phase. The only difference is the shape of the line that separates the
two Higgs phases.  For $N_c=2$, the model with $v=0$ is invariant
under a larger global symmetry group, the ${\rm Sp}(N_f)/{\mathbb
  Z}_2$ group~\cite{BPV-19}. In this case, the line $v=0$, which is a
first-order line for $N_f\ge 3$, separates the Higgs phases. For $N_c
> 2$ instead, there is no additional symmetry and the boundary of the
two Higgs phases is a generic curve that lies in the positive $v$
region, see Fig.~\ref{phdianfgtncge3}.

In the following we focus on the SU(2)-gauge NAH theory
(\ref{lattNAH}), which should be already fully representative for the
problem we address in this paper.  We recall that the analysis of the
RG flow of the NAH field theory, see Sec.~\ref{sft}, indicates that
the attraction domain of the stable charged FP must be located in the
region $v>0$.  Therefore, we should focus on the continuous
transitions occurring in the domain $v>0$, where the symmetry-breaking
pattern is~\cite{BFPV-21-su}
\begin{equation}
{\rm U}(N_f)\rightarrow \mathrm{SU}(2)\otimes\mathrm{U}(N_f-2).
\label{glsymbrpat}
\end{equation}

\section{Numerical analyses of the multiflavor lattice SU(2) NAH models}
\label{numres}

The numerical results reported in Ref.~\cite{BFPV-21-su} for the SU(2)
lattice gauge model provided good evidence of continuous transitions
for $v=1$, $\gamma=1$, and $N_f=40$.  First-order transitions were
instead observed for $N_f=20$, for several values of $\gamma$ and
$v$. Therefore, a natural hypothesis is that for $v=1$ and $\gamma=1$
(more generally, for generic positive $v$ and sufficiently large
values of $\gamma$) the transitions are continuous for $N_f>N_f^*$,
with $N_f^*$ in the interval $20<N_f^*<40$.

To understand whether these transitions are associated with the
charged FP of the NAH field theory, we need accurate numerical results
that can be compared with predictions obtained from the 3D NAH field
theory. We will focus on the critical exponent $\nu$, comparing the
numerical estimates with the large-$N_f$ result, Eq.~(\ref{nulargen}).
For this purpose, we have performed numerical simulations for $v=1$,
$\gamma=1$, and $N_f=30$, $N_f=40$, and $N_f=60$, varying $\beta$
across the transition line. Simulations have been performed on cubic
lattices with periodic boundary conditions. 

The update of the link variables has been performed by using an overrelaxation algorithm combining 
standard heat-bath \cite{KP-85} and microcanonical \cite{Creutz-87} updates for SU(2)
gauge theory. A similar strategy has been adopted for the complex scalar field
$\Phi^{af}$, using a combination of two different update schemes: the first one
is a Metropolis update \cite{MRRTT-53}, which rotates two randomly chosen
elements of $\Phi_{\bm x}^{af}$, denoted by $\phi_1$ and $\phi_2$ in
the following. More precisely the trial state is of the form
\begin{equation}
\begin{aligned}
\phi'_1 &= \cos\theta_1 e^{i\theta_2} \phi_1 + \sin\theta_1 e^{i\theta_3}\phi_2\\
\phi'_2 &= 
   -\sin\theta_1 e^{i\theta_2}\phi_1 + \cos\theta_1 e^{i\theta_3}\phi_2 \,,
\end{aligned}
\label{metro_rotation}
\end{equation}
where the angles $\theta_i$ are uniformly distributed in $[-\alpha,\alpha]$ (to
ensure detailed balance), and the value of $\alpha$ has been chosen for the
acceptance probability of the update to be roughly 30\%. This update
has been supplemented by a second update scheme, in which the proposed state is
\begin{equation}
  \Phi'_{\bm x} = \frac{2 \mathrm{Re}\Tr(\Phi^\dagger_{\bm x} S_{\bm x})}
       {\Tr( S^\dagger_{\bm x} S_{\bm x})}S_{\bm x} 
- \Phi_{\bm x}\,,
\label{generalized-overrelaxed}
\end{equation}
where $S_{\bm x}$ is the matrix
\begin{equation}
    S_{\bm x} = \sum_\mu 
    (U_{{\bm x}, \mu} \Phi_{{\bm x}+\hat{\mu}} +
    U^\dagger_{{\bm x}-\hat{\mu}, \mu} \Phi_{{\bm x}-\hat{\mu}}) \,.
\end{equation}
It is immediate to verify that this transformation is involutive, hence it can
be used as the trial selection step of a Metropolis update. Note that for $v=0$
the transformation in Eq.~\eqref{generalized-overrelaxed} would leave the
action invariant, corresponding to a microcanonical update for the scalar
fields, hence we call pseudo-microcanonical this update scheme.  For $v>0$ the
trial state $\Phi'_{\bm x}$ is accepted or rejected using a Metropolis test,
which turns out to have a large acceptance rate for all the cases studied in
this paper.

We call lattice iteration a sequence of 11 update sweeps on all the lattice
sites, for both gauge and scalar variables. For the link variables an heat-bath
update is followed by ten microcanonical updates, while for scalars a
Metropolis update with trial state Eq.~\eqref{metro_rotation} is followed by
ten pseudo-microcanonical updates. The ratio 1 to 10 of ergodic to
microcanonical steps was kept fixed in all the cases, since autocorrelation
times turned out to be small enough for our purposes, without requiring further
parameter optimizations.  Measures where performed after every lattice
iteration, and the gathered statistic has been at least of the order of $10^5$
iterations in all the cases.  Data have been analyzed by using standard
jackknife and blocking techniques, with the maximum block size needed to
correctly estimate statistical errors never exceeding $5\times 10^{2}$.

\subsection{Observables and finite-size scaling}
\label{obsfss}

To study the breaking of the global SU($N_f$) symmetry, we monitor
correlation functions of the gauge-invariant bilinear operator
\begin{equation}
  Q_{\bm x}^{fg} = \sum_a {\bar\Phi}_{\bm x}^{af} \Phi_{\bm x}^{ag}
  - {1\over N_f} \delta^{fg}.
\label{qdef}
\end{equation}
We define its two-point correlation function (since we use periodic
boundary conditions, translation invariance holds)
\begin{equation}
G({\bm x}-{\bm y}) = \langle {\rm Tr}\, Q_{\bm x} Q_{\bm y} \rangle,
\label{gxyp}
\end{equation}
the corresponding susceptibility $\chi$, and second-moment
correlation length $\xi$ defined as 
\begin{eqnarray}
  \chi=\sum_{\bm x} G({\bm x}),\quad \xi^2 = {1\over 4 \sin^2 (\pi/L)}
           {\widetilde{G}({\bm 0}) - \widetilde{G}({\bm p}_m)\over
             \widetilde{G}({\bm p}_m)},
\label{xidefpb}
\end{eqnarray}
where ${\bm p}_m = (2\pi/L,0,0)$ and $\widetilde{G}({\bm
  p})=\sum_{{\bm x}} e^{i{\bm p}\cdot {\bm x}} G({\bm x})$ is the
Fourier transform of $G({\bm x})$.  In our numerical study we also
consider the Binder parameter
\begin{equation}
U = \frac{\langle \mu_2^2\rangle}{\langle \mu_2 \rangle^2}, \qquad
\mu_2 = L^{-6} \sum_{{\bm x},{\bm y}} {\rm Tr}\,Q_{\bm x} Q_{\bm y},
\label{binderdef}
\end{equation}
and the ratio
\begin{equation}\label{rxidef}
R_{\xi}=\xi/L.
\end{equation}

At a continuous phase transition, any RG invariant ratio $R$, such as
the Binder parameter $U$ or the ratio $R_\xi$, scales as~\cite{PV-02}
\begin{eqnarray}
  R(\beta,L) = {\cal R}(X) +  L^{-\omega} {\cal R}_\omega(X) +
  \ldots, \label{scalbeh}
\end{eqnarray}
where 
\begin{equation}
  X = (\beta-\beta_c)L^{1/\nu},
  \label{Xdef}
  \end{equation}
$\nu$ is the critical correlation-length exponent, $\omega>0$ is the
leading scaling-correction exponent associated with the first
irrelevant operator, and the dots indicate further negligible
subleading contributions.  The function ${\cal R}(X)$ is universal up
to a normalization of its argument. Also ${\cal R}_\omega(X)$ is
universal apart from a multiplicative factor and normalization of the
argument [the same of $\mathcal{R}(X)$].  In particular, $R^*\equiv
{\cal R}(0)$ is universal, depending only on the boundary conditions
and aspect ratio of the lattice.  Since $R_\xi$ defined in
Eq.~\eqref{rxidef} is an increasing function of $\beta$, we can
combine the RG predictions for $U$ and $R_\xi$ to obtain
\begin{equation}
  U(\beta,L) = {\cal U}(R_\xi) + O(L^{-\omega}),
\label{uvsrxi}
\end{equation}
where ${\cal U}$ now depends on the universality class, boundary
conditions, and lattice shape, without any nonuniversal multiplicative
factor. Eq.~\eqref{uvsrxi} is particularly convenient because it
allows one to test universality-class predictions without requiring a
tuning of nonuniversal parameters.

Analogously, in the FSS limit the susceptibility defined in
Eq.~(\ref{xidefpb}) scales as
\begin{eqnarray}
  \chi \approx L^{2-\eta_Q} {\cal C}(R_\xi),
  \label{chiscal}
  \end{eqnarray}
where $\eta_Q$ is the critical exponent that parametrizes the
power-law divergence of the two-point function (\ref{gxyp}) at
criticality, and ${\cal C}$ is a universal function apart from a
multiplicative factor.

\subsection{Numerical results}
\label{resnc2nfgt2}

We now present the FSS analyses of the observables introduced in
Sec.~\ref{obsfss}, for the SU(2) gauge theory. We set $v=\gamma=1$ and
consider $N_f=30,\,40,\,60$. The MC data are obtained for lattice
sizes up to $L=48$ for $N_f=40$ and $N_f=60$, and up to $L=42$ for
$N_f=30$. As we shall see, they are sufficient to accurately determine
the critical behavior of the lattice SU(2)-gauge NAH models
(\ref{lattNAH}).

\begin{figure}[tbp]
 \includegraphics[width=0.95\columnwidth,
   clip]{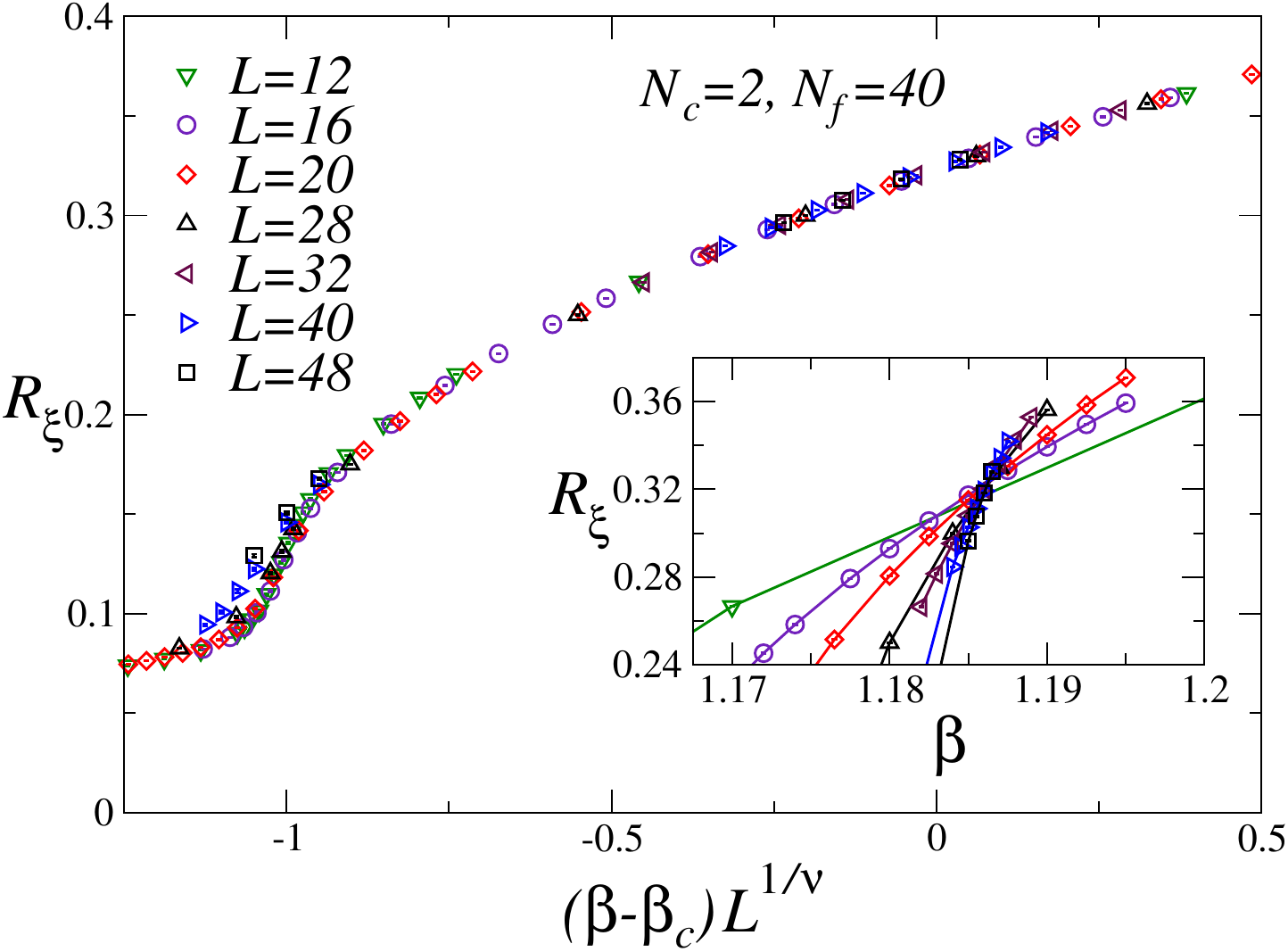}
\caption{Plot of the RG invariant ratio $R_\xi\equiv \xi/L$ versus $X = (\beta
- \beta_c) L^{1/\nu}$ for $N_f=40$, $v=1$, and $\gamma=1$, using the best
estimates $\beta_c = 1.1863$ and $\nu = 0.745$.  The data show a good scaling
behavior with increasing $L$, in particular for $X\gtrsim -1$, confirming the
asymptotic FSS behavior (\ref{scalbeh}). The inset shows the estimates of
$R_{\xi}$ versus $\beta$: fixed-$L$ data show a clear crossing point that
allows one to determine $\beta_c$.}
\label{Rxi-X-Nf40}
\end{figure}

\begin{figure}[tbp]
    \includegraphics[width=0.95\columnwidth, clip]{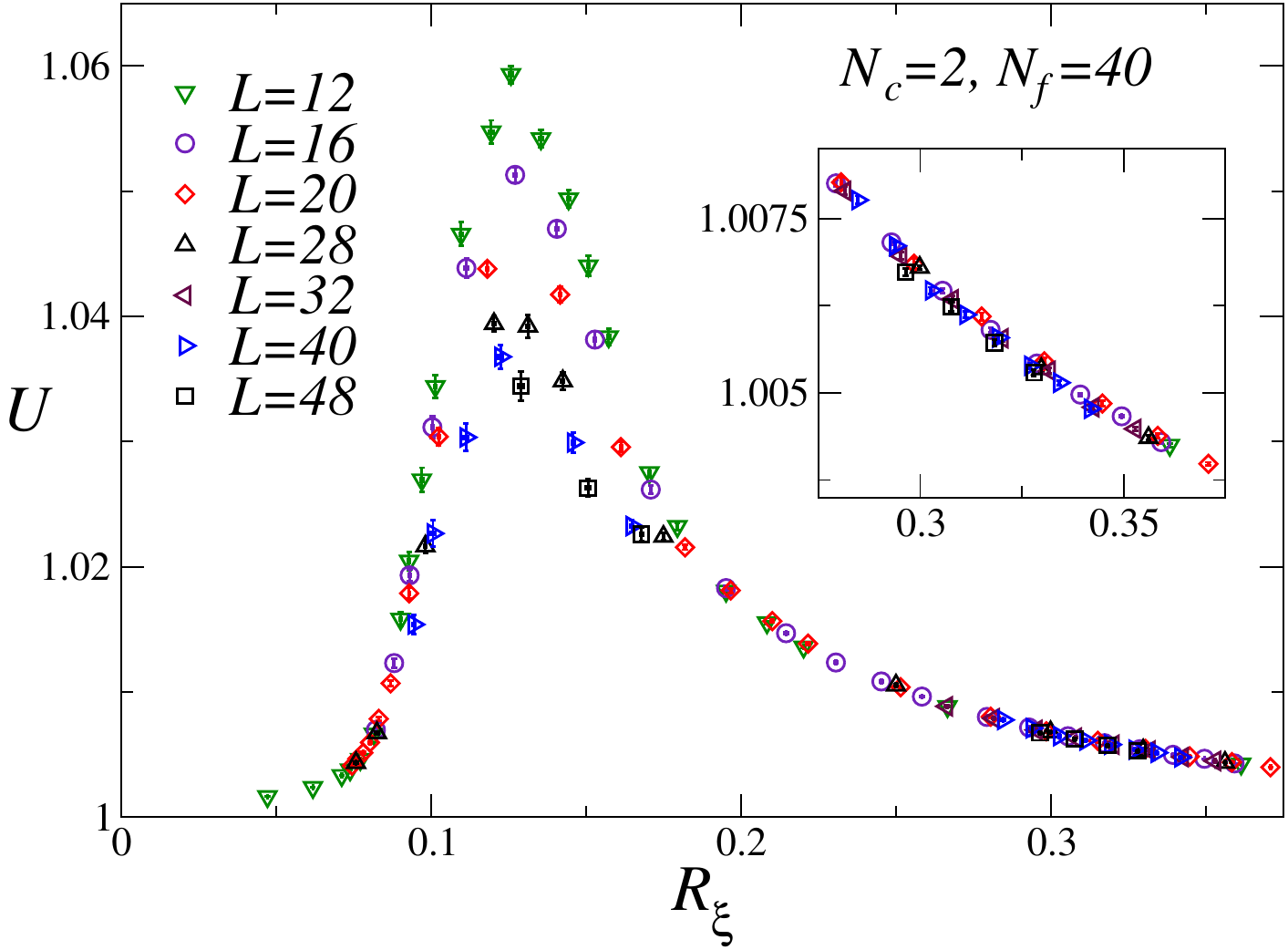}
    \caption{Binder parameter $U$ versus $R_\xi$ for $N_f=40$, $v=1$,
      and $\gamma=1$.  The data appear to converge to a scaling curve
      when increasing L, confirming the expected FSS behavior
      (\ref{uvsrxi}) characterizing a continuous transition. We also
      note that scaling corrections appear to be significantly larger
      at the peak of $U$ around $R_\xi\approx 0.12$ (corresponding to
      $X\approx -1$ in Fig.~\ref{Rxi-X-Nf40}), see also the discussion
      reported in the text.  The inset shows the same data around
      $R_{\xi}\approx 0.3$, corresponding to data around $X=0$, where
      the scaling behavior appears to be optimal, and most of the
      simulations on larger lattices have been performed.}
\label{U-Rxi-Nf40}
\end{figure}

\begin{figure}[tbp]
  \includegraphics[width=0.95\columnwidth,clip]{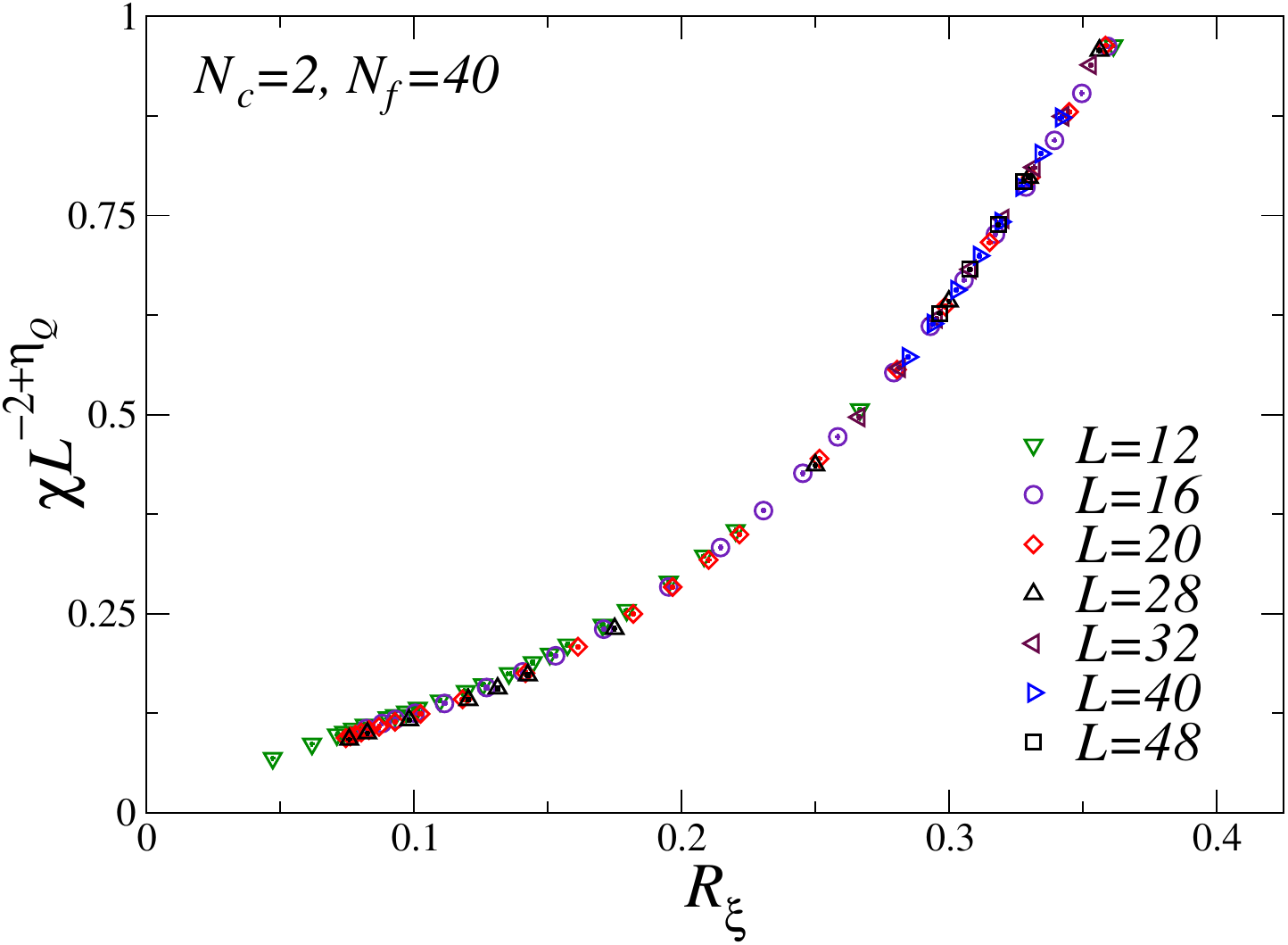}
 \caption{Ratio $\chi/L^{(2-\eta_Q)}$ versus $R_\xi$, for $N_f=40$,
   $v=1$, and $\gamma=1$, using the best estimate $\eta_Q = 0.87$. The
   collapse of the data onto a single curve is excellent, confirming
   the validity of the FSS scaling relation, Eq.~(\ref{chiscal}).}
\label{chi-X-Nf40}
\end{figure}

To begin with, we discuss the behavior for $N_f=40$, a case that was already
considered in Ref.~\cite{BFPV-21-su}. Here we consider significantly larger
systems and obtain more accurate data.  Estimates of $R_\xi$ are shown in
Fig.~\ref{Rxi-X-Nf40} for several values of $L$, up to $L=48$, Data have a
clear crossing point for $R_\xi\approx 0.32$, which indicates a transition at
$\beta \approx 1.186$.  Accurate estimates of the critical point $\beta_c$ and
of the critical exponent $\nu$ are determined by fitting $R_\xi$ to the
expected FSS behavior (\ref{scalbeh}).  We perform several fits, parametrizing
the function ${\cal R}(X)$ with an order-$n$ polynomial (stable results are
obtained for $n\gtrsim 3$) and also including $O(L^{-\omega})$ corrections with
$\omega$ in the range $[0.5,1.0]$. Note that $\omega$ is generally expected to
be smaller than one and to approach one in the large-$N_f$ limit, as in the 3D
$N$-vector models~\cite{ZJ-book}. In any case, results are almost independent
of the value of $\omega$. Moreover, to have an independent check of the role of
the scaling corrections, fits have been repeated, systematically discarding the
data for the smallest lattice sizes (i.e. including only data for $L\ge L_{\rm
min}$ with $L_{\rm min}=8,12,16,20$ typically).  Data points corresponding
to different $\beta$ values come from independent simulations, and
are thus independent of each other (no reweighting has been used). The large
number of data points ensures the stability of the fitting procedure, and
combining the results of all the fits with $\chi^2/\mathrm{dof}\lesssim 1.5$ we
obtain the estimates
\begin{equation}
  \beta_c=1.1863(1),\quad \nu=0.745(15),\quad {\rm for}\;N_f=40,
\label{estimates-nuNf40}
\end{equation}
where the errors also take into account how the results change when the
fit parameters are varied in reasonable ranges, which is the main source
of systematical error (these results are compatible within errors with results
reported in Ref.~\cite{BFPV-21-su} using smaller lattice sizes, up to $L=28$).
In Fig.~\ref{Rxi-X-Nf40} we plot $R_\xi$ versus $X=(\beta-\beta_c)L^{1/\nu}$
using the above estimates of $\beta_c$ and $\nu$.  The resulting scaling
behavior when increasing $L$ definitely confirms the correctness of the
estimates reported in Eq.~(\ref{estimates-nuNf40}). Some sizeable scaling
corrections are observed only for $R_{\xi}\lesssim 0.12$, corresponding to
$X\lesssim -1$, however the convergence of large lattices, $L\gtrsim 30$ say,
is clear also in that region (see also the following discussion on scaling corrections
present in Fig.~\ref{U-Rxi-Nf40}). We also mention that consistent, but less
precise, results are obtained by analyzing the Binder parameter $U$.

Further evidence of FSS is achieved by the unbiased plot of the Binder
parameter $U$ versus $R_\xi$, cf. Eq.~(\ref{uvsrxi}), see
Fig.~\ref{U-Rxi-Nf40}. Again we observe a nice scaling behavior for
$R_\xi\gtrsim 0.2$, see in particular the inset of
Fig.~\ref{U-Rxi-Nf40} where data around $R_\xi\approx 0.3$ are shown.
We also note that sizable scaling corrections are observed around the
peak of $U$, corresponding to $R_\xi\approx 0.12$, which is also the
region where the scaling behavior of $R_\xi$ versus $X$ show larger
scaling corrections.  These corrections are consistent with the
expected $L^{-\omega}$ asymptotic approach and $\omega\approx 1$, in
particular the values of $U$ at the peak turn out to behave as $U_{\rm
  peak}\approx a + b/L^\omega$ with $\omega \approx 1$. It is also
important to note that, although significant corrections are present
in the peak region, the peak values decrease when increasing the
lattice size, excluding a discontinuous transition (if the transition
were of first order, the Binder parameter would diverge for
$L\to\infty$ \cite{CLB-86,VRSB-93,CPPV-04}).

We have also estimated the exponent $\eta_Q$ characterizing the
behavior of the susceptibility $\chi$.  Using the expected FSS
behavior (\ref{chiscal}), $\eta_Q$ was estimated by fitting $\log
\chi$ to $(2-\eta_Q) \log L + C(R_\xi)$, using a polynomial
parametrization for the function $C(x)$. Proceeding as in the analysis
of $R_\xi$, we obtain $\eta_Q = 0.87(1)$.  The resulting FSS plot is
shown in Fig.~\ref{chi-X-Nf40}.

\begin{figure}[tbp]
    \includegraphics[width=0.95\columnwidth, clip]{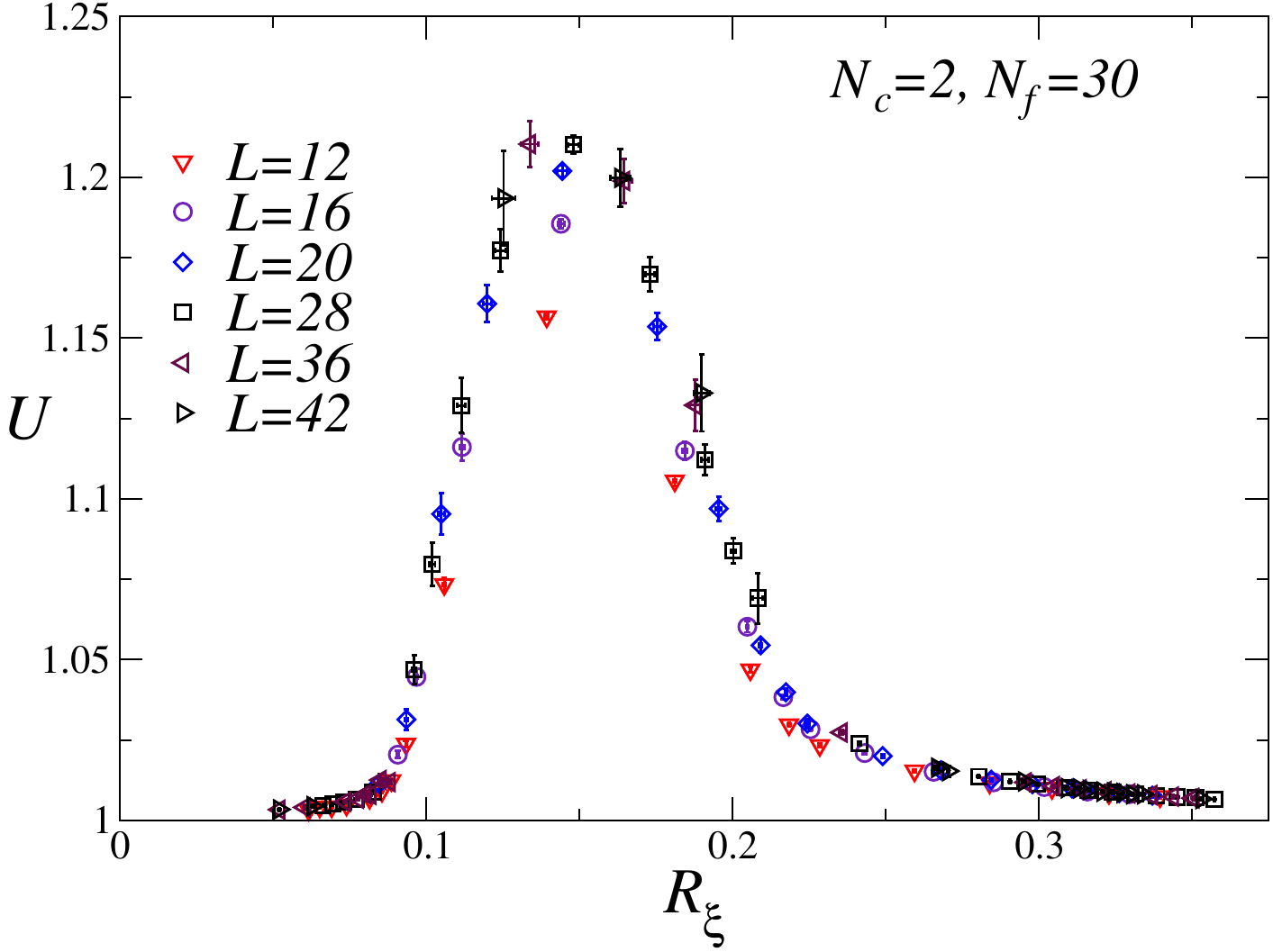}
    \caption{Binder parameter $U$ versus the ratio $R_\xi$ for
      $N_f=30$, $v=1$, and $\gamma=1$. Data converges to a scaling
      curve with increasing $L$, in agreement with Eq.~(\ref{uvsrxi}),
      with some small deviations, which can easily explained by the
      presence of power-law suppressed scaling corrections.}
\label{U-Rxi-Nf30}
\end{figure}

\begin{figure}[tbp]
 \includegraphics[width=0.95\columnwidth, clip]{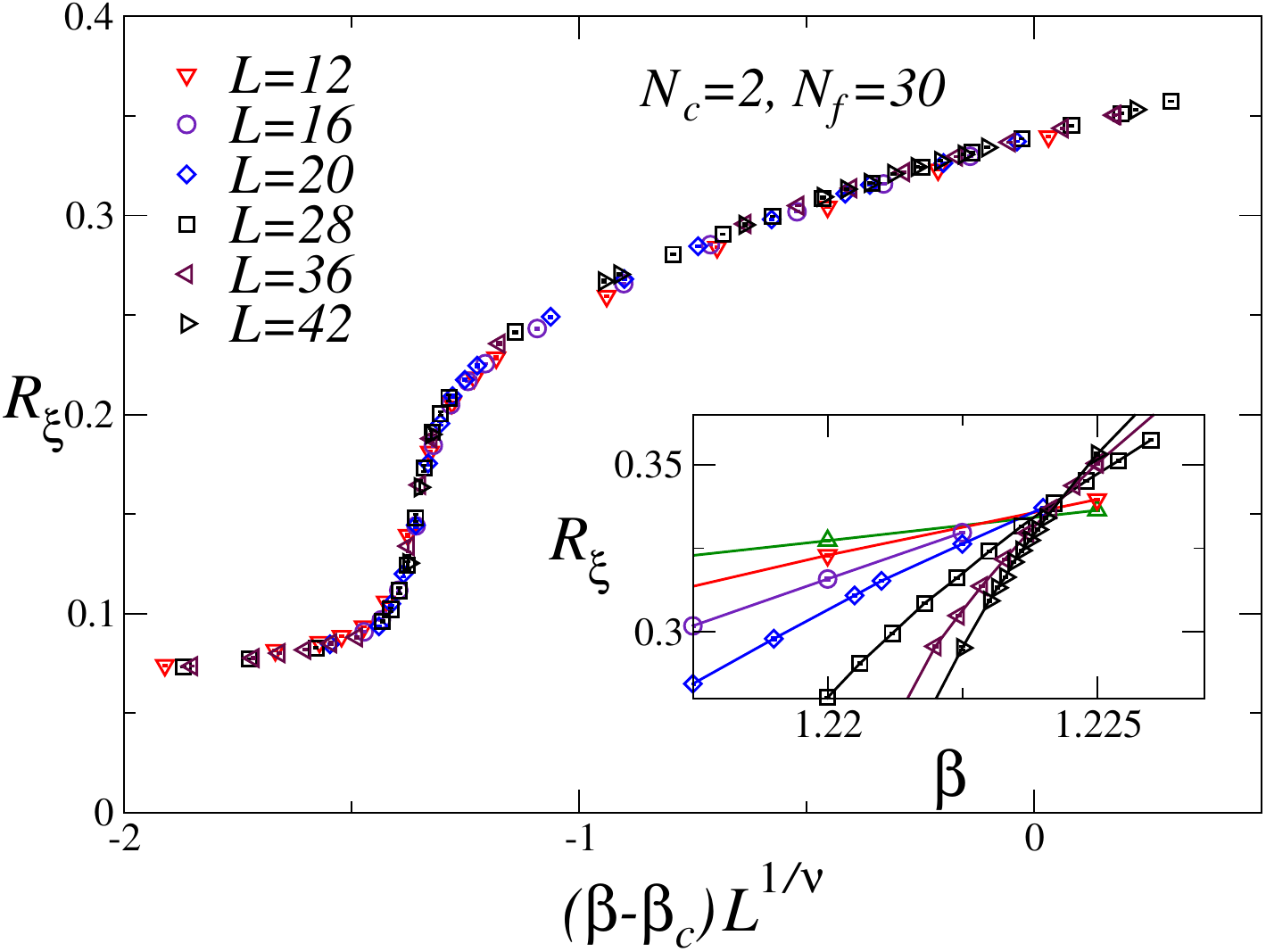}
 \caption{Plot of the RG invariant ratio $R_\xi\equiv \xi/L$ versus $X
   = (\beta - \beta_c) L^{1/\nu}$ for $N_f=30$, $v=1$, and $\gamma=1$,
   using the best estimates $\beta_c = 1.22435$ and $\nu = 0.64$. The
   good scaling of the data nicely confirms the asymptotic FSS
   behavior (\ref{scalbeh}).  The inset reports the estimates of
   $R_\xi$, showing a crossing at the critical point $\beta_c$, versus
   $\beta$, for $R_\xi\approx 0.335$.}
\label{URxi-X-Nf30}
\end{figure}

The MC data obtained for $N_f=30$ and $N_f=60$ (again for $v=1$ and
$\gamma=1$) can be analyzed analogously. In both cases we observe a
clear evidence of a continuous transition.  In particular, the Binder
parameter $U$ approaches an asymptotic FSS curve when plotted versus
$R_\xi$, see, e.g., Fig.~\ref{U-Rxi-Nf30}.  By fitting $R_\xi$ to the
FSS ansatz (\ref{scalbeh}), analogously to the case $N_f=40$, we
obtain the estimates
\begin{equation}
  \beta_c=1.22435(10),\quad \nu=0.64(2),\quad {\rm for}\;N_f=30,
\label{estimates-nuNf30}
\end{equation}
and
\begin{equation}
  \beta_c=1.1416(1),\quad \nu=0.81(2),\quad {\rm for}\;N_f=60,
\label{estimates-nuNf60}
\end{equation}
where again the errors take into account the small variations of the
results when changing the fit parameters. A FSS plot of $R_\xi$ for
$N_f=30$ is shown in Fig.~\ref{URxi-X-Nf30}.  We have also estimated
the exponent $\eta_Q$.  Performing the same analysis of the
susceptibility as for $N_f=40$, we obtain the estimates
$\eta_Q=0.79(1)$ for $N_f=30$ and $\eta_Q=0.910(5)$ for $N_f=60$.

\begin{figure}[tbp]
   \includegraphics[width=0.95\columnwidth, clip]{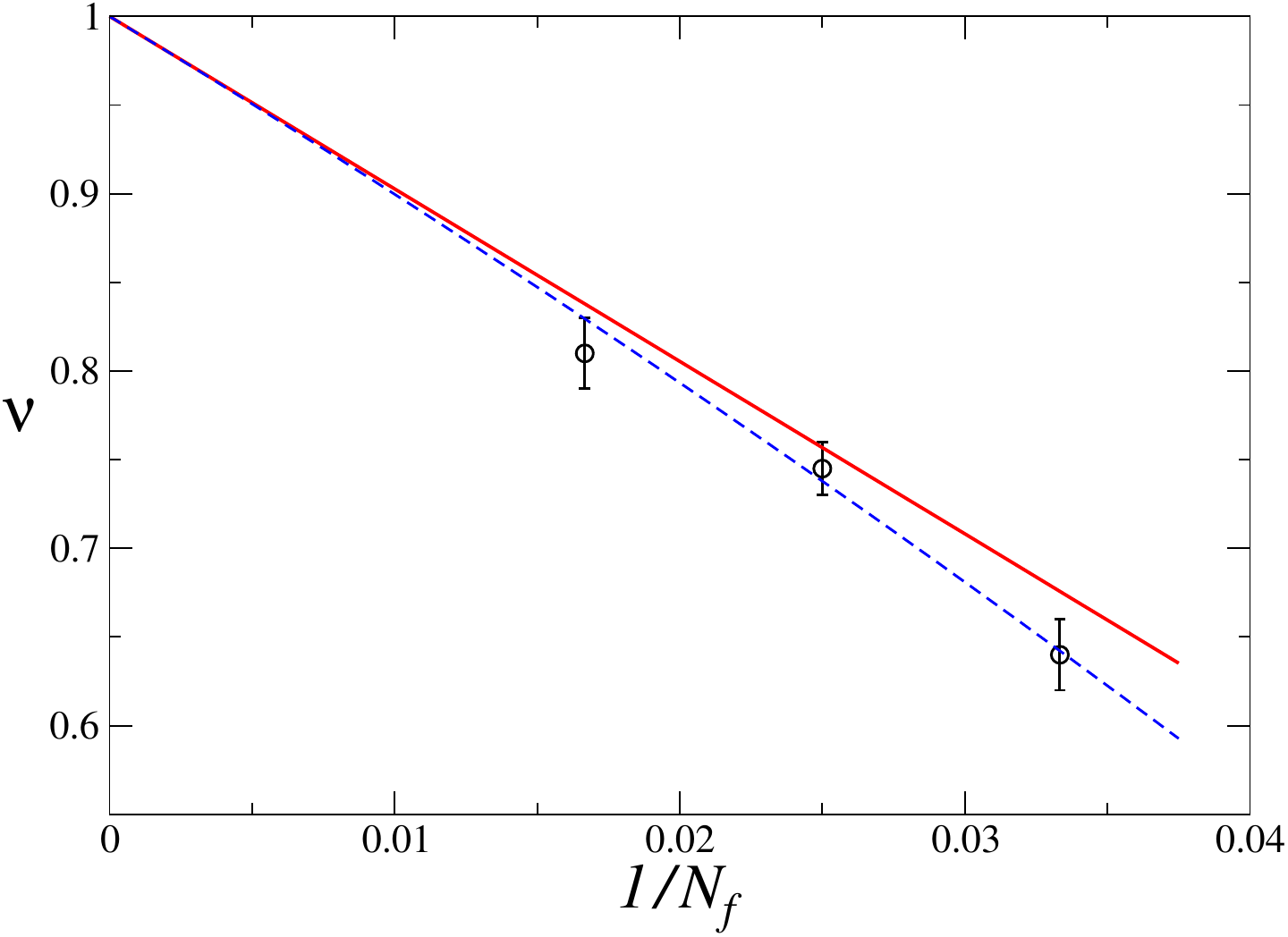}
   \caption{MC estimates of the critical exponent $\nu$ versus
     $1/N_f$.  For comparison we also report the $O(1/N_f)$
     theoretical prediction, Eq.~(\ref{nulargen}) (solid line), and a
     next-to-leading interpolation $\nu=1-9.727/N_f+a/N_f^2$ (dashed
     line), obtaining $a=-30(10)$.  }
\label{largencomp}
\end{figure}

We now compare the above results for $\nu$ with the large-$N_f$
prediction, Eq.~(\ref{nulargen}), see Fig.~\ref{largencomp}. The
agreement is satisfactory, For instance, Eq.~(\ref{nulargen}) predicts
$\nu=0.757$ for $N_f=40$ and $N_c=2$, to be compared with the MC
result $\nu=0.745(15)$.  Concerning the exponent $\eta_Q$, the
numerical estimates are compatible with the limiting value $\eta_Q=1$
for $N_f\to\infty$, which generally holds for any bilinear operator of
scalar fields. Finite-$N_f$ results are consistent with a $1/N_f$
correction, as expected. A fit of the data gives $\eta_Q\approx 1 -
c/N_f$ with $c\approx 5$ for $N_f\gtrsim 40$.

The nice agreement between the numerical estimates of $\nu$ and 
the field-theoretical large-$N_f$ prediction allows us to conclude 
that, for $\gamma > 0$ and $v>0$ and large values of $N_f$, 
transitions along the line that separates the disordered from the 
Higgs phase are continuous and naturally associated
with the charged FP of the SU(2)-gauge NAH field theory
(\ref{cogau}).  
We expect this result to hold also for larger values of $N_c$.

\section{Conclusions}
\label{conclu}

We consider 3D lattice SU($N_c$) gauge Higgs models with U($N_f$)
global invariance with the purpose of identifying continuous
transition lines with a critical behavior associated with the stable
charged FP of the RG flow of the NAH field theory defined by the
Lagrangian (\ref{cogau}).  This would imply that the lattice models
admit a continuum limit that provides a nonperturbative definition of
the NAH field theory, as it occurs for lattice QCD~\cite{Wilson-74}.

We focus on SU(2) gauge theories. We present results of MC simulations
for a relatively large number of flavors, in order to be able to
compare the MC results with field-theoretical $1/N_f$ predictions.
The RG flow of the SU(2)-gauge NAH field theory has a stable charged
FP in the region $v>0$, for $N_f > N^*_f$. Close to four dimensions,
$N_f^*$ is very large, $N^*_f \approx 376$, see
Sec.~\ref{sft}. However, our 3D numerical results show that continuous
transitions in the relevant parameter region occur for significantly
smaller numbers of components.  While for $N_f=20$ only first-order
transitions (for different values of $v$ and $\gamma$) are
observed~\cite{BFPV-21-su}, for $N_f=30$ a continuous transition is
found for $v=\gamma=1$. These results suggest that $20<N_f^*<30$, or
equivalently that $N_f^*=25(4)$ in three dimensions. More importantly,
the numerical estimates of the length-scale critical exponent $\nu$
for $N_f=30,40,60$ are in nice agreement with the large-$N_f$
field-theoretical result, Eq.~(\ref{nulargen}).  As far as we know,
this is the first evidence of the existence of critical behaviors in
3D lattice NAH models that can be associated with the charged FP of
the 3D SU($N_c$)-gauge NAH field theory.

As we mentioned in Sec.~\ref{sft-relevance} not all transitions in gauge
systems require an effective description in terms of a gauge field theory.
There are many instances in which gauge fields have no role.  In these cases
the effective model is a scalar LGW theory in which the fundamental field is a
(coarse-grained) gauge-invarianct scalar order parameter.  This approach was
employed in Refs.~\cite{PW-84,BPV-03,PV-13} to discuss the nature of the
finite-temperature transition of QCD in the chiral limit. Indeed, it was
assumed that the transition was only due to the condensation of a
gauge-invariant operator, bilinear in the fermionic fields.  Such operator was
then taken as fundamental field in an effective 3D LGW $\Phi^4$ theory, whose
RG flow was supposed to determine the nature of the chiral transition.  The
implicit assumption was that only gauge-invariant fermionic related modes are
relevant critical modes.

It is thus worth discussing the predictions of the LGW approach in the
present case, to exclude that the transitions that we have discussed
above have an effective LGW description. In the LGW approach the
fundamental field is a hermitian traceless $N_f\times N_f$ matrix
field $\Psi({\bm x})$, which represents a coarse-grained version of
the gauge-invariant bilinear operator $Q_{\bm x}$ defined in
Eq.~(\ref{qdef}). The corresponding most general LGW Lagrangian with
global SU($N_f$) symmetry is~\cite{PV-19,BPV-19}
\begin{eqnarray}
&&{\cal L}_{\rm LGW} = \hbox{Tr }\partial_\mu \Psi \partial_\mu \Psi +
  r\, \hbox{Tr} \Psi^2
\label{SLGW} \\ 
&& \qquad + \,w\, \hbox{Tr}\,\Psi^3 + 
      u\, (\hbox{Tr}\,\Psi^2)^2 + v \,\hbox{Tr}\,\Psi^4.
\nonumber
\end{eqnarray}
For $N_f=2$ the cubic term vanishes and the two quartic terms are
equivalent.  In this case a continuous transition is possible in the
SU(2)/${\mathbb Z}_2$, that is in the O(3) vector, universality class.
For $N_f > 2$ the cubic term is present and, on the basis of the usual
mean-field arguments, one expects a first-order transition also in
three dimensions (unless a tuning of the model parameters is performed
to cancel the cubic term).  Therefore, the LGW approach does not give
the correct predictions for the transitions that we have investigated.
The reason of the failure is likely related to the fact that the LGW
approach assumes that gauge fields are not relevant at criticality. In
LGW transitions their only role is that of restricting the critical
modes to the gauge-invariant sector. Instead, the relation between the
critical transitions we observed and the NAH field theory implies that
gauge fields are critical and relevant for the critical behavior in
the cases we studied.

We should note that the results presented here are valid for $v > 0$.  For $v <
0$ continuous transitions are observed for $N_f=2$, in the O(3) universality
class~\cite{BFPV-21-su}.  The NAH field theory does not provide their correct
effective description, since there are no stable FPs in the RG flow of the NAH
field theory with negative $v$ for any $N_f$.  On the other hand, the LGW
theory predicts O(3) transitions for $N_f=2$, since the Lagrangian (\ref{SLGW})
is equivalent to the O(3) Lagrangian for this value of $N_f$.  We conclude
that, for $v < 0$ and $N_f=2$, gauge modes do not play any role and the
transition admits a LGW description.

This discussion shows that the critical behavior of 3D models (or 4D models at
finite temperature) with non-Abelian gauge symmetry is quite complex  and
possibily more interesting than expected.  In particular, the knowledge of the
order parameter of the transition is not enough to characterize the critical
behavior.  Informations on the behavior of the gauge fields are required to
identify the correct effective description.

\acknowledgments

The authors acknowledge support from project PRIN 2022 ``Emerging
gauge theories: critical properties and quantum dynamics''
(20227JZKWP).  Numerical simulations have been performed on the CSN4
cluster of the Scientific Computing Center at INFN-PISA.

\end{document}